\documentclass[12pt]{article}
\title{Randomized opinion dynamics over networks:\\ influence estimation from partial observations}
\author{Chiara Ravazzi, Sarah Hojjatinia, Constantino M. Lagoa, \\Fabrizio Dabbene% <-this % stops a space
\thanks{Chiara Ravazzi and Fabrizio Dabbene are with the Institute of Electronics, Computer and Telecommunication Engineering, National Research Council of Italy (CNR-IEIIT), c/o Politecnico di Torino, Corso Duca Degli Abruzzi, 10129, Italy. 
E-mail: chiara.ravazzi@ieiit.cnr.it, fabrizio.dabbene@ieiit.cnr.it.\newline
\indent Sarah Hojjatinia and Constantino M. Lagoa are with the Department of Electrical Engineering at the Pennsylvania State University, University Park, PA 16802, USA. 
E-mail: szh199@psu.edu, lagoa@psu.edu.\newline
\indent
The authors' research has been partially supported by the International Bilateral Joint CNR Lab COOPS.}%
}
\date{}
\usepackage{graphicx}
\usepackage{amssymb,amsmath,dsfont}
\usepackage{color}
\usepackage{pgfplots}
\usepackage{subfigure}
\usepackage{tikz}
\usetikzlibrary{external}
\usepackage[title]{appendix}
\newcommand{\real}{\mathbb{R}}

\newcommand{\integernonnegative}{\mathbb{Z}_{\ge 0}}

\newcommand{\R}{\mathbb{R}} % used for real
\newcommand{\G}{\mathcal{G}} % graph
\newcommand{\V}{\mathcal{V}} % node set
\newcommand{\E}{\mathcal{E}} % edge set

\newcommand{\card}[1]{|#1|}  	% cardinality

\newcommand{\Exp}{\mathds{E}} % expectation
\def\Exp{\mathbb{E}}

\newcommand{\1}{\mathds{1}} % vector of ones
\newcommand{\diag}{\operatorname{diag}} % diagonal of a matrix

\newcommand{\argmin}[1]{\underset{#1}{\mathrm{argmin\,}}}
\newtheorem{example}{Example}

\newtheorem{theorem}{Theorem}
\newtheorem{lemma}{Lemma}

\newtheorem{proposition}{Proposition}

\usepackage{tikz}
\usepackage{subfigure}

\begin{document}
\maketitle
\thispagestyle{empty}
\pagestyle{empty}
\begin{abstract}
In this paper, we propose a technique for the estimation of the influence  matrix in a sparse social network, in which $n$ individual communicate 
in a gossip way. At each step, a random subset of the social actors is active and interacts with randomly chosen neighbors. The opinions evolve according to a Friedkin and Johnsen mechanism, in which the individuals updates their belief to a convex combination of their current belief, the belief of the agents they interact with, and their initial belief, or prejudice. 
Leveraging recent results of estimation of vector autoregressive processes, we reconstruct the social network topology and the strength of the interconnections starting from \textit{partial observations} of the interactions, thus removing one of the main drawbacks of finite horizon techniques.
The effectiveness of the proposed method is shown on  randomly generation networks.
 
\end{abstract}
\section{Introduction}
Recent years have witnessed the growth of a new research direction, at the boundary between social sciences and control theory, interested in studying dynamical social networks (DSN).
 As pointed out in the recent survey \cite{ProTempo:2017-1}, ``this trend was enabled by the introduction of new mathematical models describing dynamics of social groups, the advancement in complex networks theory and multi-agent systems, and the development of modern computational tools for big data analysis."
Aim of this line of research is to study the mechanisms underlying \textit{opinion formation}, that is to analyze how the individuals' opinions are modified and evolve as a consequence of the interactions among different agents connected together through a network of relationships.

To this end, several models have been proposed in the literature based on different communication mechanisms, see again \cite{ProTempo:2017-1} for a nice survey.
These models were proven to be able to explain certain behaviors observed in the evolution of opinions, such as the emergence of consensus as in French-De-Groot models~\cite{French_1956,Degroot_74,Chatterjee1977}, or the persistent disagreement in social systems when stubborn agents are present \cite{NEF-ECJ:99,Hegselmann.Krause2002OpinionDynamicsand,Acemoglu:2013:OFD:2448396.2448397}.
Among these models, special attention has received the Friedkin and Johnsen's (F\&J) model \cite{FRIEDKINJOHNSEN1990}, which has been experimentally validated for small and medium size groups \cite{NEF-ECJ:99,FriedkinDec2011}. In the F\&J model, the agents are influenced by the others' opinions, but are not completely open-minded, being persistently driven by their initial opinions. More precisely, at each round of communication the agents update their beliefs by taking a convex combination of the opinions coming from the neighbors, weighted with respect a influence matrix, and their prejudices. It should be noted that this model extends the French-DeGroot model with stubborn agents, which is included  as a particular case. 

A common characteristic of most of the models presented in the literature is to assume synchronous interactions: in iterative rounds, all individuals interact with their neighbors (adjacent nodes in the relationship graph), and their opinions are updated taking into account others' opinion and their relative \textit{influence} (strength of the interactions). Recently, the somehow unrealistic assumption of simultaneous interactions has been lifted, by introducing a version of F\&J model  in which the interactions occur following a \textit{gossip} paradigm \cite{FrascaTempo:2015}. In particular, it was shown that the average opinion still exhibits the salient convergence properties of the original synchronous model.

The study of DSN has been very active in this years. On the one side, the theoretical properties of the proposed models have been thoroughly investigated \cite{ProTempo:2017-1}, on the other side, these model have been used to  detect communities \cite{Fortunato20161} and to define new centrality measures to identify social leaders~\cite{citeulike:392803,PhysRevE.70.056131}. These latter have opened the way to the research on control of DSN, intended as the ability of placing influencers in an optimal way~\cite{Yildiz:2013:BOD:2542174.2538508,liu11}.

However, most research in this field is based on the assumption that \textit{the social network is given}, and in particular the influences among individuals are known. For instance, in the experiments performed by Friedkin and Johnsen in~\cite{FriedkinJohnsen:1999}, the relative interpersonal influence was ``measured" by introducing a mechanism in which actors were asked to distribute ``chips" between the actors they interacted with. 
Clearly, these ad-hoc solutions are not viable in case of large networks.  At the same time, however, 
we have now the availability of a large amount of data, especially in the case of online social networks, and tools for measuring in real-time the individuals opinions are becoming available
(as e.g.\ ``likes" on Facebook or Twitter, instant pools, etc.).

These considerations led to a new research direction, less explored so far, which aims at directly identifying the influence network based on collected observations. This idea is outlined  in \cite{journals/tsipn/WaiSL16}, which proposes a method to infer the network structure in French-De-Groot models with stubborn agents.

Two different approaches, known as {\em finite-horizon} and {\em infinite
horizon} identification procedures, can be adopted~\cite{Parsegov2017TAC}.
In the finite-horizon approach the opinions are observed for~$T$ subsequent rounds of conversation. Then, if enough observations are available,  the  influence matrix is estimated as the matrix best fitting the dynamics for $0 \le k \le T$, by employing classical identification techniques. This method however requires the knowledge of the discrete-time indices for the observations made, and the storage of the whole trajectory of the system. The loss of data from one agents in general requires to restart the experiment. More importantly, the updates usually occur at an unknown interaction rate, and the interaction times between agents are not observable in most scenarios, as in \cite{Wang2015103}.

The infinite-horizon identification procedure performs the estimation based on the observations of the initial and final opinions' profile only, hence it is applicable only to stable models. This techniques was proposed in \cite{RavazziTempoDabbene:2018} to estimate the influence matrix of a F\&J model, under the assumption that the matrix is sparse (i.e. agents are influenced by few friends). In particular, by using tools from compressed sensing theory, under suitable assumptions, theoretical conditions to guarantee that the estimation problem is well posed and sufficient requirements on the number of observation ensuring perfect recovery are derived.

While the infinite horizon approach is surely innovative under various aspects, it suffers the drawback of being \textit{static}. Indeed, the identification does not exploit the dynamical nature of the system, and it requires knowledge of initial and final  opinion on several different topics to build up the necessary information to render the problem identifiable. Even if \cite{RavazziTempoDabbene:2018} shows that number of topics that is strictly smaller than the size of the graph, and in many cases scales logarithmically with it, this information may be sometimes hard to collect

The present work presents a solution that goes in the direction of overcoming the main difficulties of both approaches: we propose a technique which exploits the dynamical evolution of the opinions, but at the same time does not require perfect knowledge of the interaction times, and it can be adapted to cases when some information is missing or partial.
The main idea is to make recourse to tools recently developed in the context of identification of vector autoregressive (VAR) processes~\cite{Yonina-1,Yonina-2}.

\section{Problem formulation}
\subsection{F\&J opinion dynamics with random interactions}
We consider a finite population $\V$ of interacting social actors (or agents), whose {\it social network} of potential interactions is represented by an undirected graph $\G=(\V,\E)$. At time $k\in\integernonnegative$, each agent $i\in\V$ holds a {\it belief or opinion} about a specific subject under discussion. We denote the vector of beliefs as $x(k)\in\real^\V$. We have $(i,j)\in \E$ if $j$ may directly influence the opinion of agent $i$. To prevent trivialities, we assume that $\card{\V}>2$.\
\ Let $W\in \real^{\V\times\V}$ be a nonnegative matrix which describs the strength of the interactions ($W_{ij}=0$ if $(i,j)\not\in \E$) and $\Lambda$ be a diagonal matrix defining how sensitive each agent is to the opinions of the others based on interpersonal influences. We assume that $W$ is row-stochastic, i.e., $W\1=\1$ and we set $\Lambda=\diag(\lambda)$. We denote the set of neighbors of node $i\in\V$ by the notation $\mathcal{N}_i =\{j\in\V : (i,j)\in\E\}$ and the degree $d_i =|\mathcal{N}_i|$.

The dynamics evolves as follows. 
Each agent $i\in \V $ possesses an initial belief $x_i(0)=u_i\in \real$. At each time  $k\in\integernonnegative$ a subset of nodes $\V_k$ is randomly selected from a uniform distribution over $\V$. If the node $i$ is selected at time $k$, agent $i$ interacts with a randomly chosen neighbor $j$ and updates its belief to a convex combination of its previous belief, the belief of $j$, and its initial belief. Namely, for all $i\in\V_k$
\begin{align}
\label{eq:gossip-friedkin}
\nonumber x_i(k+1)&=\lambda_{i}\big((1-W _{ij})x_i(k)+W _{ij}x_j(k)\big)+(1-\lambda_{i})u_i\\
x_\ell(k+1)&=x_\ell(k)\qquad \forall \ell\in \V\setminus\V_k,
\end{align}
 
The dynamics (2) can be formally rewritten in the following form: given $\V_k$ and $\theta(k)=\{\theta_i\}_{i\in\V_k}$, we have
$$x(k+1)=A(k)x(k)
+B(k)u,
$$
where
$$
A(k)=(I -\sum_{i\in\V_k}e_ie_i^{\top} (I - \Lambda)) (I + \sum_{i\in\V_k}W_{i\theta_i} (e_ie_{\theta_i}^{\top}  -e_ie_i^{\top} ) ),
$$
$\theta_i=j$ with probability $1/d_i$, 
and
$$
B(k)= \sum_{i\in\V_k}e_ie_i^{\top}(I-\Lambda).
$$

It can be shown that, due to the random nature of the dynamical system and to the pairwise interactions, the opinion dynamics fails to converge in a deterministic sense and shows persistent oscillations. However, under suitable conditions we can guarantee the convergence of the expected dynamics and the ergodicity of the oscillations. More precisely, we have the following two results, whose proof is a direct consequence of the results in~\cite{FrascaTempo:2015}.
\vskip 2mm

\begin{proposition}\label{prop:ex_dyn}
Assume that in the graph
associated to $ W$ for any node $\ell\in \V$ there exists a path from
$\ell$ to a node $i$ such that  $\lambda_{i}<1$. Then
$$
\mathbb{E}[x(k+1)]=\overline{A} \mathbb{E}[x(k)]+\overline b
$$
with 
\begin{gather*}
\overline A=\left(1-\beta\right)I+\beta\Lambda(I- D^{-1}(I-W))
\\
\overline{b}=\beta(I-\Lambda)u
\end{gather*}
$\beta=|\V_k|/|\V|$ and $D$ is the degree matrix of the network, a diagonal matrix whose
diagonal entry is equal to the degree $d_i = |\mathcal{N}_i|$.
Moreover the sequence $\mathbb{E}[x(k)]$ converges  to 
\[
\mathbb{E}[x(\infty)]=(I-\overline{A})^{-1}\overline{b}.
\]
\end{proposition} 
\vskip 2mm

\begin{theorem}[Ergodic opinion dynamics]
\label{thm:gossip-opinions} 
Assume that in the graph
associated to $W$ for any node $\ell\in \V$ there exists a path from
$\ell$ to a node $i$ such that  $\lambda_{i}<1$.
Then, the dynamics~\eqref{eq:gossip-friedkin} is ergodic, and the time-averaged opinions  
$\overline{x}(k)=\frac{1}{k}\sum_{\ell\leq k} x(\ell)$ converge to $\mathbb{E}[x(\infty)].$
\end{theorem}
\vskip 2mm

We illustrate these properties through the following example.

\begin{example}We consider the Zachary's Karate Club dataset extracted from~\cite{Zac77}.
%(see the database in https://...), 
%whose graph is depicted in Figure~\ref{fig:KC_net}. 
This graph represents the friendships between 34 members of a karate club at a US university in the 1970s. 
The number of connections is equal to $|\mathcal{E}| = 78$ and the maximal degree of the network is $17$. The nonzero entries of influence matrix $W$ are generated according uniform distribution in the range $[0,1]$ and then the rows are normalized to make $W$ row-stochastic. The sensitive parameters $\lambda_i$ are extracted from a uniform distribution in the range of $[0.9,1]$. 

\begin{figure}
\subfigure[Evolution of the opinions.]{
\centering\includegraphics[width=0.9\columnwidth]{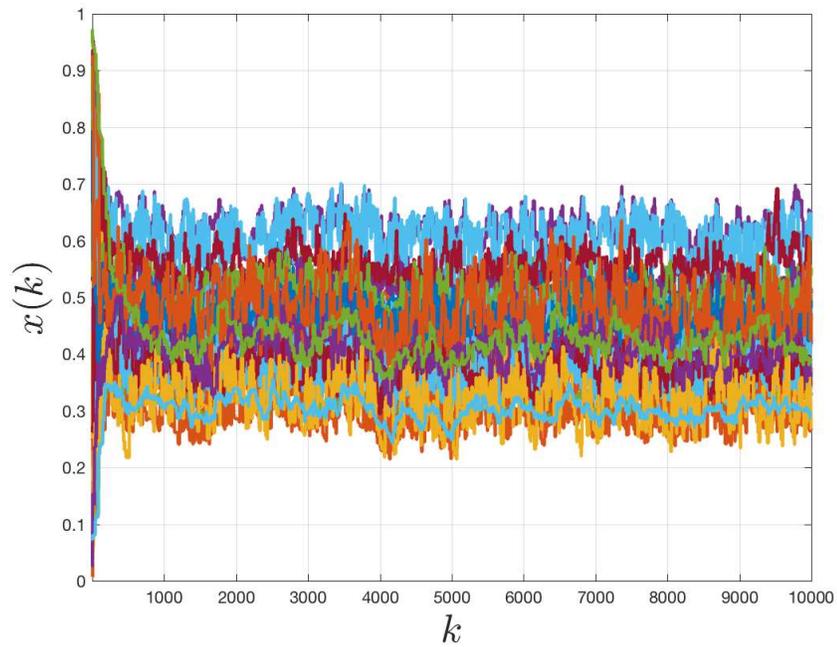}\label{fig:fig0}
}
\subfigure[Evolution of the time averages.]{
\centering\includegraphics[width=0.9\columnwidth]{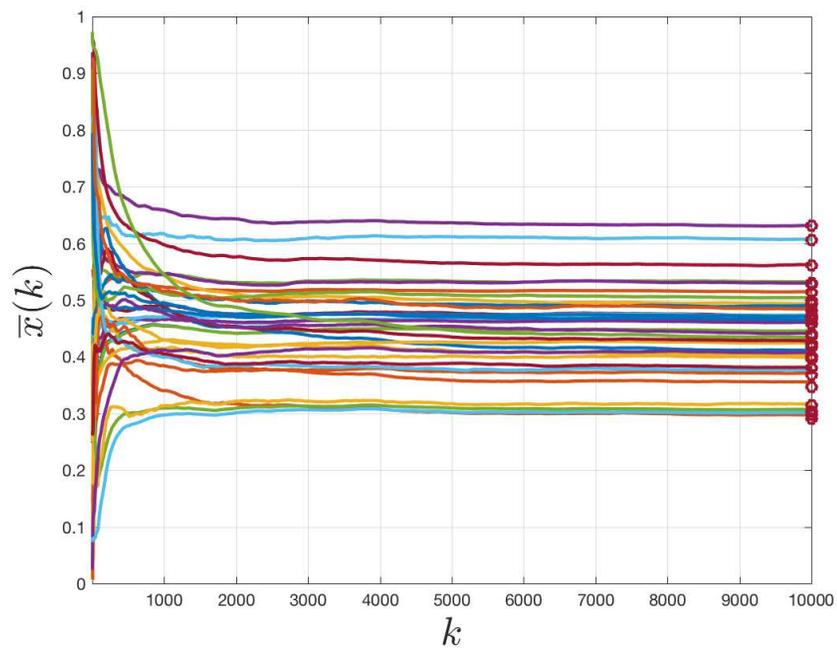}\label{fig:fig1}
}
\caption{Friedkin and Johnsen dynamics with random interactions in Zachary karate club network}\label{fig:FJ_ZK_net}
\end{figure}
In Figure \ref{fig:fig0} it is shown the evolution of the opinions. It should be noticed that the dynamics does not converge and the oscillations are persistent over the time. However, these fluctuations are ergodic and the time averages converge to a limit point corresponding to the limit point of the expected dynamics (see red circles in Figure \ref{fig:fig1}).
\end{example}
\subsection{Observation models}
Let us consider the opinion dynamics in \eqref{eq:gossip-friedkin} where the influence matrix $W$ is unknown. At each time we observe partial information of the form
\begin{equation}\label{eq:obs}
z(k)=P(k)x(k)
\end{equation}
where $P(k)$ is a random measurement matrix defined by
$$
P(k)=\mathrm{diag}(p(k))
$$
and $p(k)\in\{0,1\}^{\V}$ is a random selection vector with known distribution.
In particular, in this setting the following observation schemes will be considered:

\subsubsection{Intermittent observations}
If we let
\[
p(k)=\begin{cases}\1&\text{w.p.}\ \rho\\
0&\text{otherwise}
\end{cases}
\]
then at $k\in\integernonnegative$ all observations are available with probability $\rho$ or no observations at all are observed.
This model allows to capture the typical situation in which the actual rates at which the interactions occur is not perfectly known, 
and thus sampling time is different from interaction time.

\medskip
\subsubsection{Independent random sampling}
At each time $k\in\integernonnegative$ the selection vector is $p_i(k)\sim \mathrm{Ber}(\rho_i)$ for all $i\in\V$, that is the opinions are observed independently with probability $\rho_i\in[0,1]$. We refer the specific case in which the observations are made with equal probability $\rho_i=\rho$ for all $i\in\V$ to {\em independent and homogeneous sampling}. In particular, if $\rho= 1$ we have full observations, and if $\rho\neq 1$, we have partial information.
This model has a clear interpretation for DSN, capturing situations in which only a subset of individuals can be contacted (e.g.\ random interviews).

Given the sequence of observation $\{z(k)\}_{k=1}^{t}$ we are interested in the estimation of the matrix $W$, call it $\widehat{W}_t$, and derive theoretical conditions on the number of samples that are sufficient to have an error not larger than a fixed tolerance $\epsilon$ with high probability.
\section{Influence estimation}
\subsection{Proposed algorithm}
To reconstruct the influence matrix, we start by introducing the opinions' cross-correlation matrix, which is defined as follows
$$
\Sigma^{[\ell]}(k):=\mathbb{E}\left[x(k)x(k+\ell)^{\top}\right].
$$
Then, we follow the identification approach proposed in \cite{Yonina-1}. To this end, we can derive the following theorem, which provides a description of the evolution of the covariance matrix  $\Sigma^{[\ell]}(k)$. The proof is reported in Appendix \ref{app:A}.

\vskip 2mm
\begin{theorem}\label{thm:YW}
Assume that in the graph
associated to $W$ for any node $\ell\in \V$ there exists a path from
$\ell$ to a node $i$ such that  $\lambda_{i}<1$.
Then for all $k\in\integernonnegative$ we have
$$
\Sigma^{[1]}(k)=\Sigma^{[0]}(k)\overline{A}^{\top}+\mathbb{E}[x(k)]\overline{b}^{\top}.
$$
Moreover $\Sigma^{[0]}(k)$ and $\Sigma^{[0]}(k)$ converge in a deterministic sense to $\Sigma^{[0]}({\infty})$ and $\Sigma^{[1]}(\infty)$, respectively, satisfying
\begin{equation}\label{eq:YW}
\Sigma^{[1]}(\infty)=\Sigma^{[0]}(\infty)\overline{A}^{\top}+\mathbb{E}[x(\infty)]\overline{b}^{\top}.
\end{equation}

\end{theorem}
\vskip 2mm

It should be noticed that the relation in \eqref{eq:YW} is a sort of Yule-Walker equation, %\cite{XX} 
used for estimation in autoregressive processes. The above equation provides some hints to identify the influence matrix $W$ by replacing the theoretical covariances $\Sigma^{[\ell]}(k)$ with estimated value $\widehat{\Sigma}^{[\ell]}(k)$. Our approach can be summarized as follows:
\begin{itemize}
\item we estimate $x(\infty)$ $\Sigma^{[0]}(\infty)$ and $\Sigma^{[1]}(\infty)$ from partial observations $\{z(k)\}_{k=1}^t$;
\item we use the provided estimations of the cross-correlation matrices to approximate $\overline{A}$ through \eqref{eq:YW};
\item we recover the influence matrix $W$ exploiting Proposition \ref{prop:ex_dyn} and Theorem \ref{thm:gossip-opinions}.
\end{itemize}
\subsection{Estimating the expected opinion profile and the cross-correlation matrices}
In order to estimate the expected opinion profile $\mathbb{E}[x(\infty)]$, we consider the time averaged observations of $z(k)$. 

\begin{proposition}\label{prop:x}The following relation holds
$$
\mathbb{E}[z(k)]=\pi\circ\mathbb{E}[x(k)]
$$
where $\pi=\mathbb{E}[p(k)]$ and $\circ$ denotes the entrywise product. 
\end{proposition}
\vskip 2mm

We estimate $\mathbb{E}[z(k)]$ using time averages
$$
\overline{z}(t)=\frac{1}{t}\sum_{k=1}^{t}z(k)
$$
from which, using relation in Proposition \ref{prop:x}, we get
\begin{equation}
\label{x_estimation}
\widehat{x}_i(t)=\frac{\overline{z}_i(t)}{\pi_i}.
\end{equation}
In order to estimate the cross correlation matrices $\Sigma^{[\ell]}(\infty)$, we consider the empirical covariance matrix of observations $z(k)$. Let us denote
$$
S^{[\ell]}(k):=\mathbb{E}[z(k)z(k+\ell)^{\top}].
$$
\begin{proposition}\label{prop:S_Sigma}The following relation holds
$$S^{[\ell]}(k)=\Pi^{[\ell]}(k)\circ\Sigma^{[\ell]}(k)$$
where $\Pi^{[\ell]}=\mathbb{E}[p(k)p(k+\ell)^{\top}]$ and $\circ$ denotes the Hadamard product. 
\end{proposition}
\vskip 2mm

Since $S^{[\ell]}(k)$ is unknown we estimate $S^{[\ell]}(k)$ using time averages
$$
\widehat{S}^{[\ell]}(t)=\frac{1}{t-\ell}\sum_{k=1}^{t-\ell}z(k)z(k+\ell)^{\top}
$$
from which, using relation in Proposition \ref{prop:S_Sigma}, we get
\begin{equation}
\label{Sigma_estimation}
\widehat{\Sigma}_{ij}^{[\ell]}(t)={\widehat{S}_{ij}^{[\ell]}(t)}/{\Pi_{ij}^{[\ell]}}.
\end{equation}

The proofs of these two results follow from basic arguments, and are omitted for brevity.
\begin{example}[Independent random sampling]
In the special case of independent and homogeneous sampling we have $\pi=\rho$
\begin{align*}
 \Pi^{[0]}&=\mathbb{P}(i,j\in\V_k)=\rho I+\rho^2(\1\1^{\top}-I) \\
\Pi^{[\ell]}&=\mathbb{P}(i\in\V_k,j\in\V_{k+\ell})=\rho^2\1\1^{\top} \quad \text{if }\ell\neq 0
\end{align*}
from which $\widehat{x}(t)={\overline{z}(t)}/{\rho}$ and
\begin{gather}
\widehat{\Sigma}_{ij}^{[\ell]}(t)=\frac{1}{\rho^2}\widehat{S}^{[\ell]}(t)-\left(\frac{1-\rho}{\rho^2}\widehat{S}^{[\ell]}(t)\circ I\right) \mathbf{1}(\ell=0)
\end{gather}
\end{example}
\begin{example}[Intermittent observations]
In the special case of intermittent observations we have $\pi=\rho$
\begin{align*}
 \Pi^{[0]}&=\rho\1\1^{\top}\quad\text{and}\quad
\Pi^{[\ell]}=\rho^2\1\1^{\top} \quad \text{if }\ell\neq 0
\end{align*}
from which $\widehat{x}(t)={\overline{z}(t)}/{\rho}$ and
$\widehat{\Sigma}_{ij}^{[\ell]}(t)=\widehat{S}^{[\ell]}(t)/{\rho^2}.
$\end{example}

\subsection{Estimating the influence matrix}
Given estimations of $\widehat{\Sigma}_{ij}^{[\ell]}$, we use them to approximate $\overline{A}$ exploiting relation in \eqref{eq:YW}. More precisely, inspired by \cite{}, we propose two methods depending on the matrix properties to recover.
\begin{enumerate}
\item Dense matrices
$$
\widehat{\overline{A}}(t)^{\top}=\widehat{\Sigma}^{[0]}(t)^{\dag}(\widehat{\Sigma}^{[1]}(t)-\overline{x}(t)\overline b^{\top})
$$
\item Sparse matrices
\end{enumerate}
\begin{align*}
&\qquad\qquad\widehat{\overline{A}}(t)^{\top}=\argmin{M\in\R^{\V\times\V}}\sum_{ij}|M_{ij}|\\
&\text{s.t. }\quad\|\widehat{\Sigma}^{[0]}(t)M-(\widehat{\Sigma}^{[1]}(t)-\overline{x}(t)\overline b^{\top})]\|_{\max}\leq \eta
\end{align*}
\section{Theoretical results}
In this section we provide a theoretical analysis on the performance of the proposed estimators.
\begin{theorem}\label{thm:performance_estimations}Let $\Delta x=\widehat{x}(t)-\mathbb{E}[x(\infty)]$ and $\Delta \Sigma^{[\ell]}(t)=	\widehat{\Sigma}^{[\ell]}(t)-\widehat{\Sigma}^{[\ell]}(\infty)$. We have the following bounds
\begin{align*}\mathbb{P}(\|\Delta x(t)\|\geq \epsilon_1)&\leq \frac{C_1}{\epsilon^2(t+1)(1-\mathrm{sr}(\overline{A}))(\pi^\star)^2}
\\
\mathbb{P}(\|\Delta \Sigma^{[\ell]}(t)\|_F\geq \epsilon_2)&\leq \frac{C_2}{\epsilon^2(t-\ell+1)(1-\mathrm{sr}(\overline{Q}))(\Pi^\star)^2}
\end{align*}
with $\Pi^\star=\min_{ij}\Pi^{[\ell]}_{ij}$ and $\pi^{\star}=\min_{i\in\V}\pi_i$.
\end{theorem}
Since
$\mathrm{sr}(\overline A)\leq 1-\beta+\beta\lambda_{\max}$,
Theorem \ref{thm:performance_estimations} guarantees that with probability at least $1-\delta$ we obtain an error in the estimation of $\mathbb{E}[x({\infty})]$ not larger than 
$\epsilon_1(\delta,t)=O\left({1}/{\pi^{\star}\sqrt{(t+1)\beta(1-\lambda_{\max})}}\right).$
In the special cases of independent and homogeneous sampling and intermittent observations we obtain that the error in the estimation of $\mathbb{E}[x({\infty})]$ and $\Sigma^{[\ell]}(\infty)$ are inversely proportional to the probability to the sampling probability and to the square of the sapling probability, respectively, being $\pi^{\star}=\rho$ and $\Pi^{\star}=\rho^2$.

The proof of Theorem \ref{thm:performance_estimations} is given in Appendix \ref{app:B}.
Following similar arguments to those in \cite{Yonina-1}, it can be proved that the error in the influence matrix is proportional to the error  
$\epsilon_2$ with probability close to 1. We omit this result for brevity. 
\section{Numerical experiments}
%{\color{magenta}[Values of $\Lambda$ and $\beta$?]}

%%%%%%%%%

%\section{Simulation Results}

In this section, we provide simulations results that illustrate the performance of the proposed algorithms for influence matrix estimation. We concentrate on the intermittent full observation case. In the following examples $\beta = 1$, and diagonal values in matrix $\Lambda$ have chosen uniformly between $0.9$ and $1$. For the sparse method, $\eta$ is considered to be a function of number of samples ($N$), and equal to ${10^{-5}}/{\sqrt{N}}$.

We start by considering the case of randomly generated networks of size $50$ with node degree equal to 3. Furthermore, it is assumed that the state is observed at every time instant; i.e.,  $\rho=1$. The averaged results for 10 simulations are provided in Figure~\ref{fig:5}. As it can be seen, there is a constant rate of convergence both for the estimates of the influence matrix $\overline{A}$ and the covariance matrix $\Sigma_1$. Moreover, since the influence matrix is sparse, the $\ell_1$ estimation of the influence matrix converges faster. Furthermore, as it can be seen in Figure~\ref{fig:5}(c), the sparse method quickly identifies the active and inactive links in the network. Jaccard index has been used to compare rate of convergence in two methods. Jaccard index $J$ of two finite sets $A$ and $B$, measures the similarity between sets and is defined as
$J(A,B)={|A \cap B|}/{|A \cup B|}
$ and in our case was used to measure the different between the real and the identified set of connections. 
As it can be seen in Figure~\ref{fig:5}(c), the sparse method  needs only about 50,000  samples to identify the edges and inactive links of the  graph.

\begin{figure}[h!]
	\subfigure[Error in estimating the covariance.]{
		\centering\includegraphics[width=0.9\columnwidth]{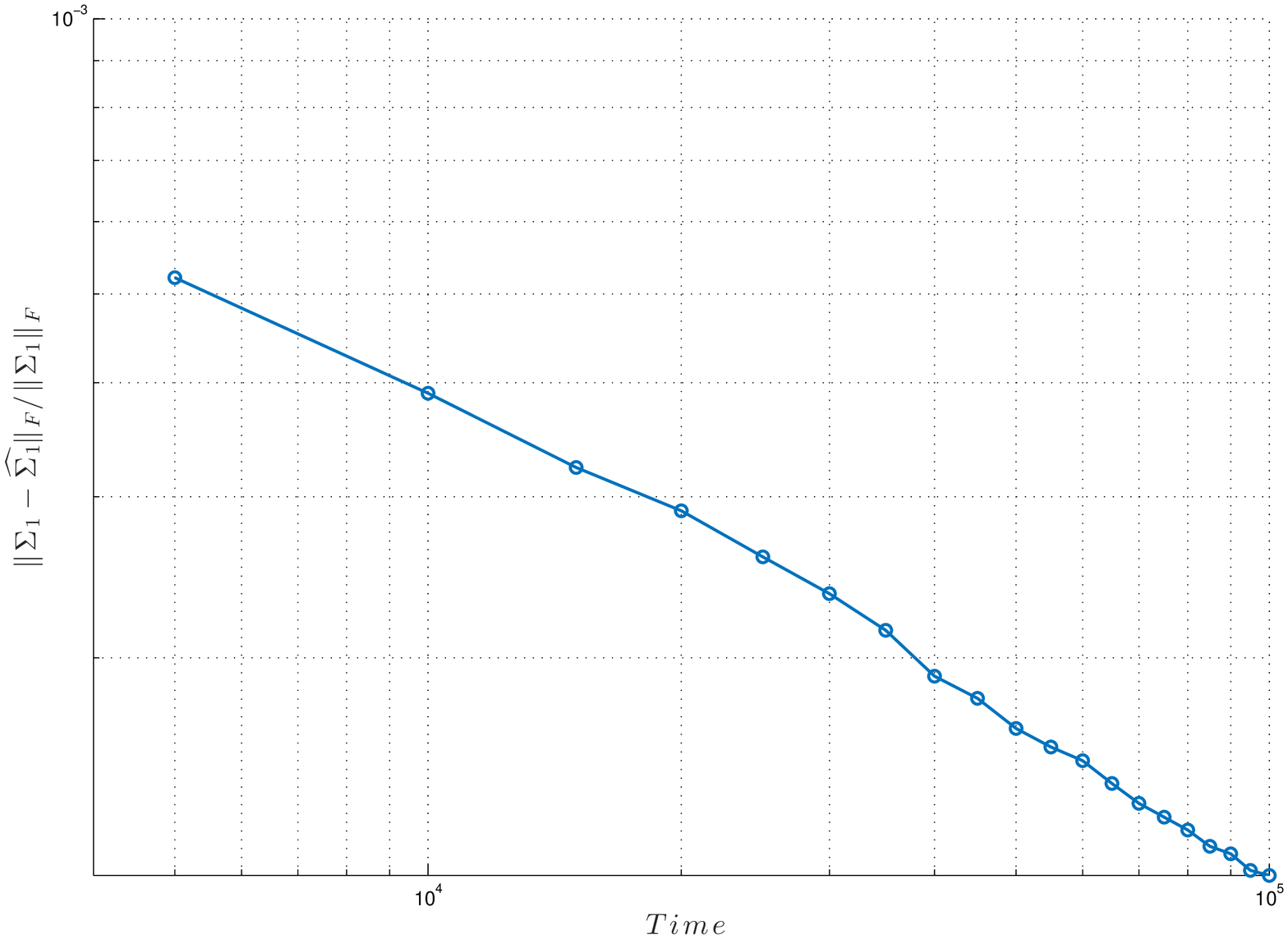} \label{fig:5b}
	}
	\subfigure[Error in estimating the influence matrix.]{
		\centering\includegraphics[width=0.9\columnwidth]{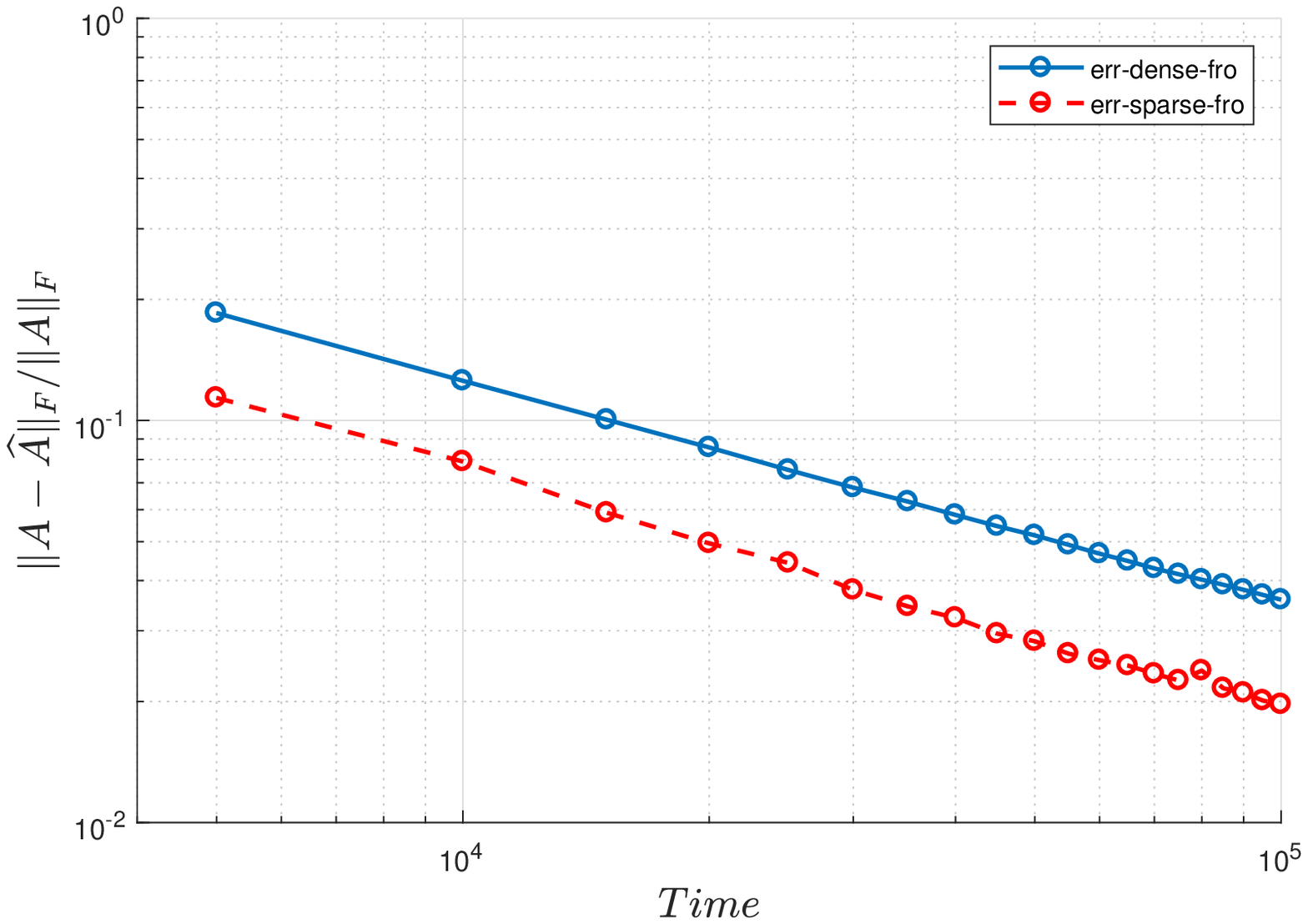} \label{fig:5a}
	}
\end{figure}
\begin{figure}[h!]
	\subfigure[Jaccard index.]{
		\centering\includegraphics[trim={5mm 0 5mm 5mm},clip,width=0.9\columnwidth]{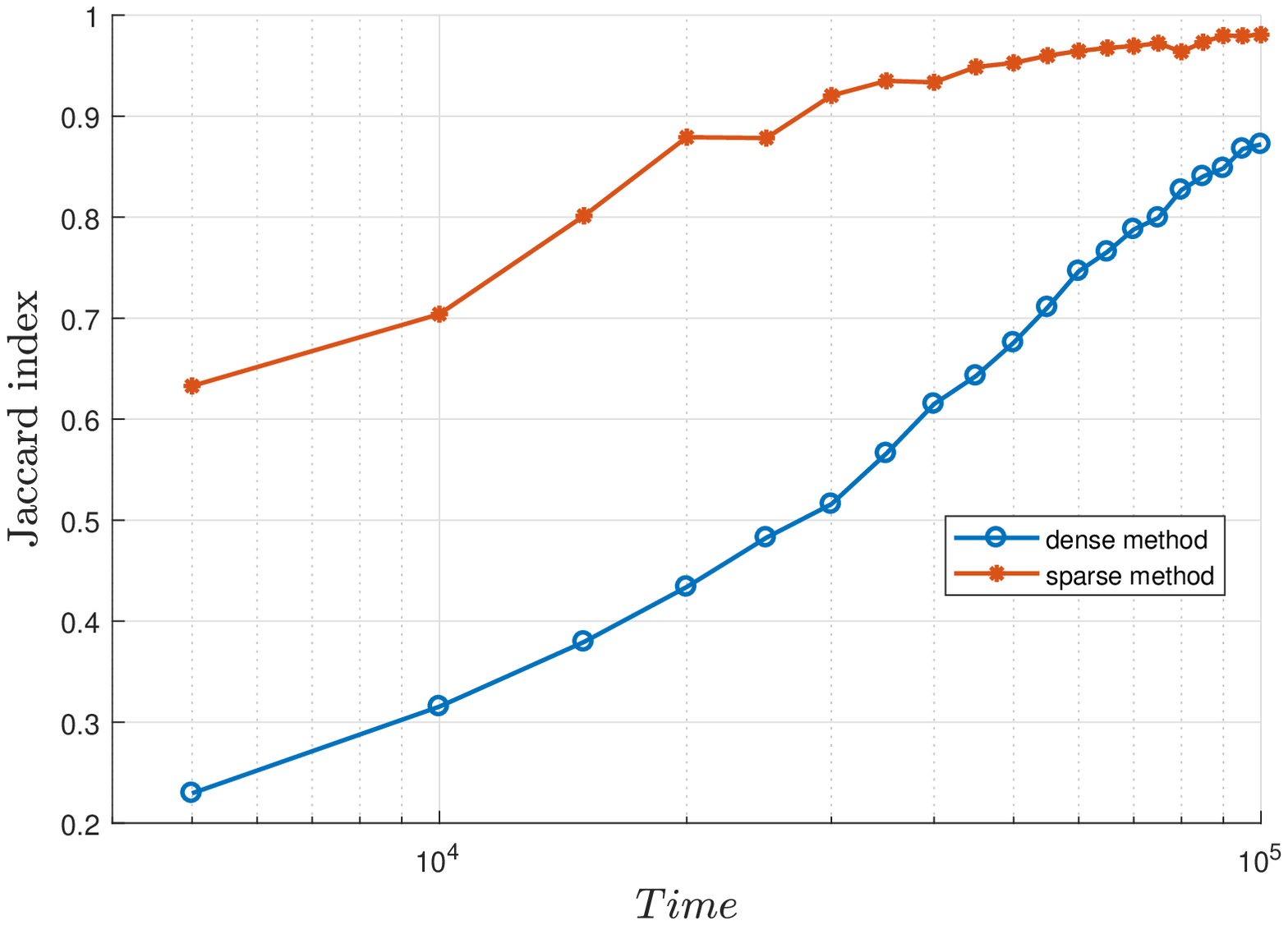} \label{fig:5c}
	}
	\caption{Comparison of dense and sparse method in a network with degree 3 and $\rho=1$}\label{fig:5}
\end{figure}

To test the scalability of the proposed method, we performed simulations for several network sizes. The averaged results for ten simulations are depicted in Figure~\ref{fig:4}. As expected, the rate of convergence is similar for all network sizes, showing that the proposed approach scales well with the dimension of the problem. Again, the method where sparsity of the connecting graph is encouraged outperforms the method that relies on inversion of the estimate of the covariance matrix.

\begin{figure}[h!]
\subfigure[Error in estimating the covariance .]{
		\centering\includegraphics[width=0.82\columnwidth]{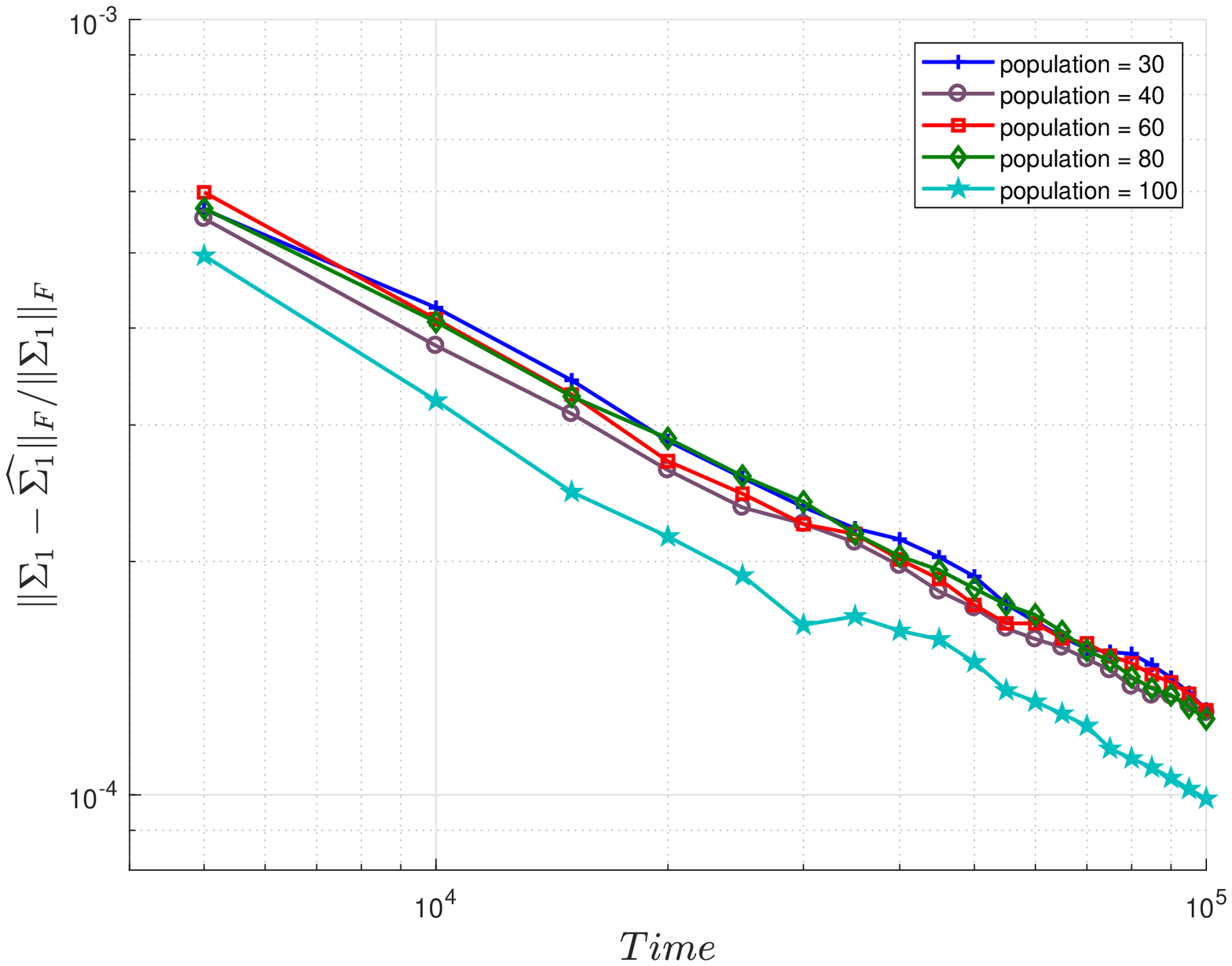} \label{fig:4c}
	}
	\subfigure[Error in estimating the influence matrix with dense method.]{
		\centering\includegraphics[width=0.82\columnwidth]{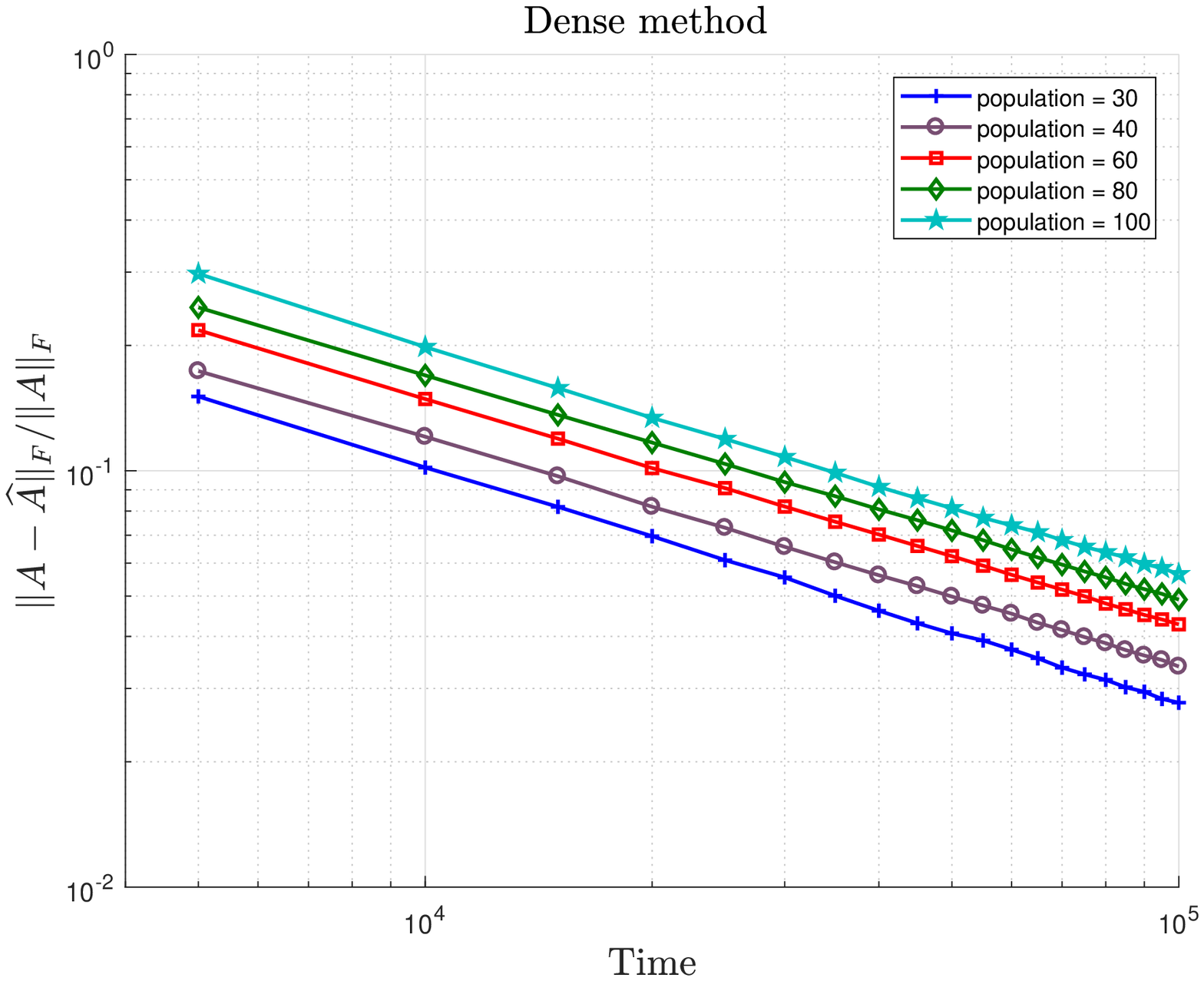} \label{fig:4a}
	}
\end{figure}
\begin{figure}[h!]
	\subfigure[Error in estimating the influence matrix with sparse method.]{
		\centering\includegraphics[width=1\columnwidth]{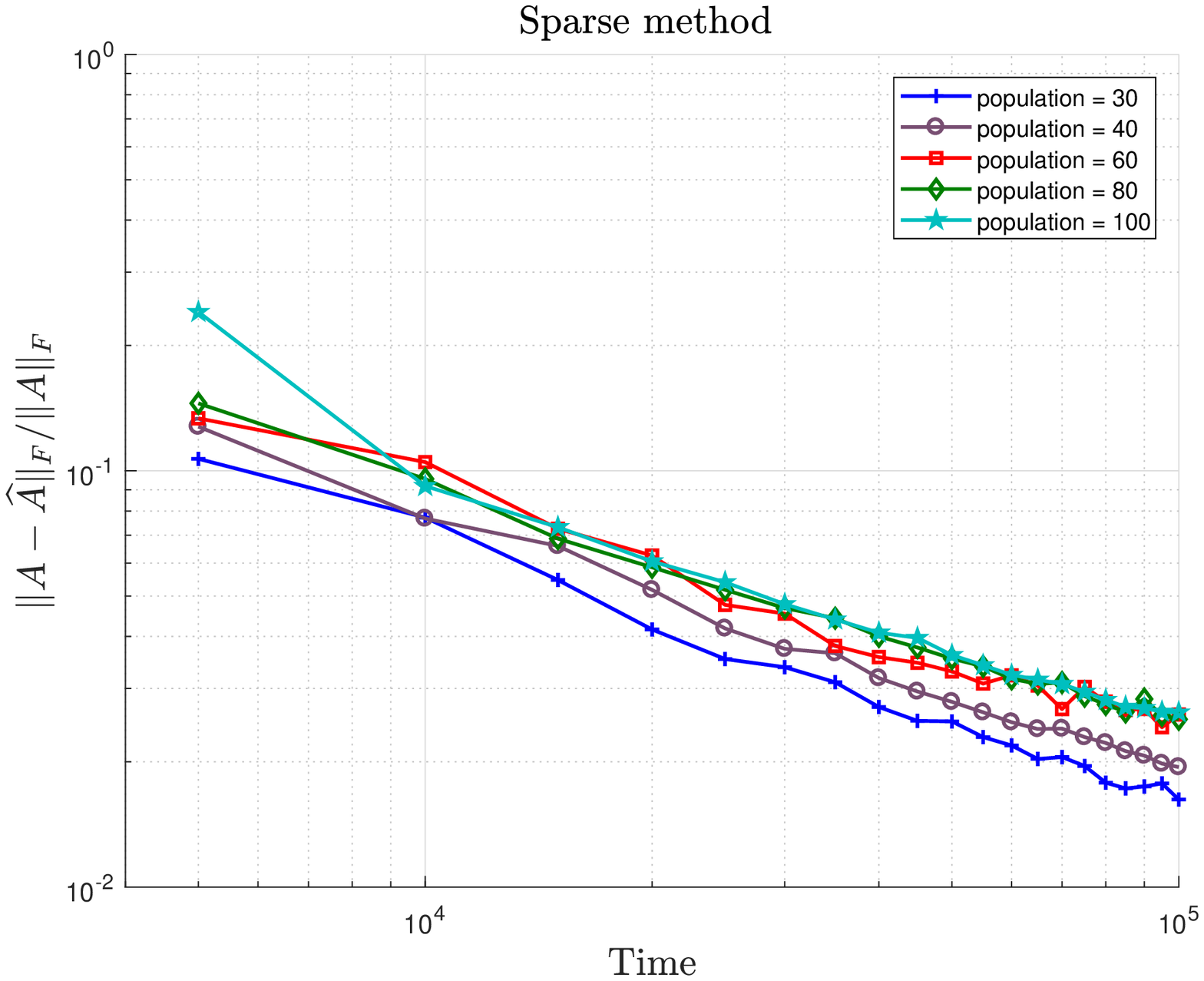} \label{fig:4b}
	}
	
	\caption{Comparison of dense and sparse method in a network with degree~3 and $\rho=1$ and different sizes of network}\label{fig:4}
\end{figure}

In the results mentioned above it was assumed that the state was measured at all time instants. To test how the algorithm performs under intermittent observations, we performed simulations for different values of the probability of observation~$\rho$. The results of 40 simulations were averaged and these are depicted in Figure~\ref{fig:3}. As expected, as the probability of observation decreases the accuracy of the estimate decreases. However, for the dense method, the rate of convergence is the same for any value of $\rho$. The impact of $\rho$ was larger in the sparse method where the performance decreased substantially as we decreased the value of $\rho$ to $0.95$. Surprisingly, smaller values of $\rho$ did not have a substantial impact on the performance of the sparse method; i.e., there is very a small impact in performance as we decrease the value of $\rho$ from $0.95$ to $0.90$, and further to $0.75$. In other words the sparse method was seen to be very resilient to changes in the probability of observation. As before, the sparse method outperformed the dense method for all values of $\rho$.

\begin{figure} [h!]
	\subfigure[Error in estimating the covariance.]{
		\centering\includegraphics[width=0.8\columnwidth]{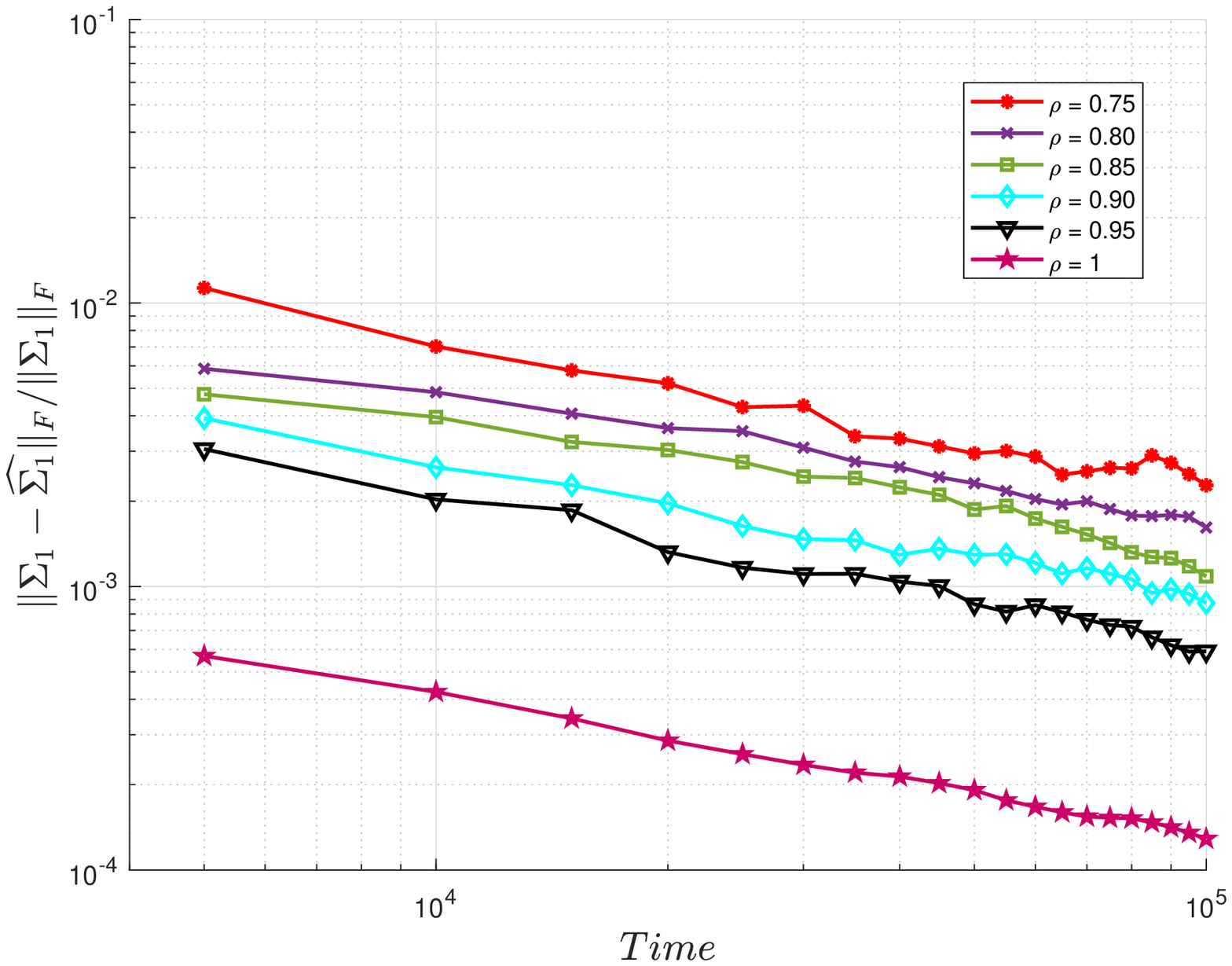} \label{fig:3c}
	}
	\subfigure[Error in estimating the influence matrix with dense method.]{
		\centering\includegraphics[width=0.8\columnwidth]{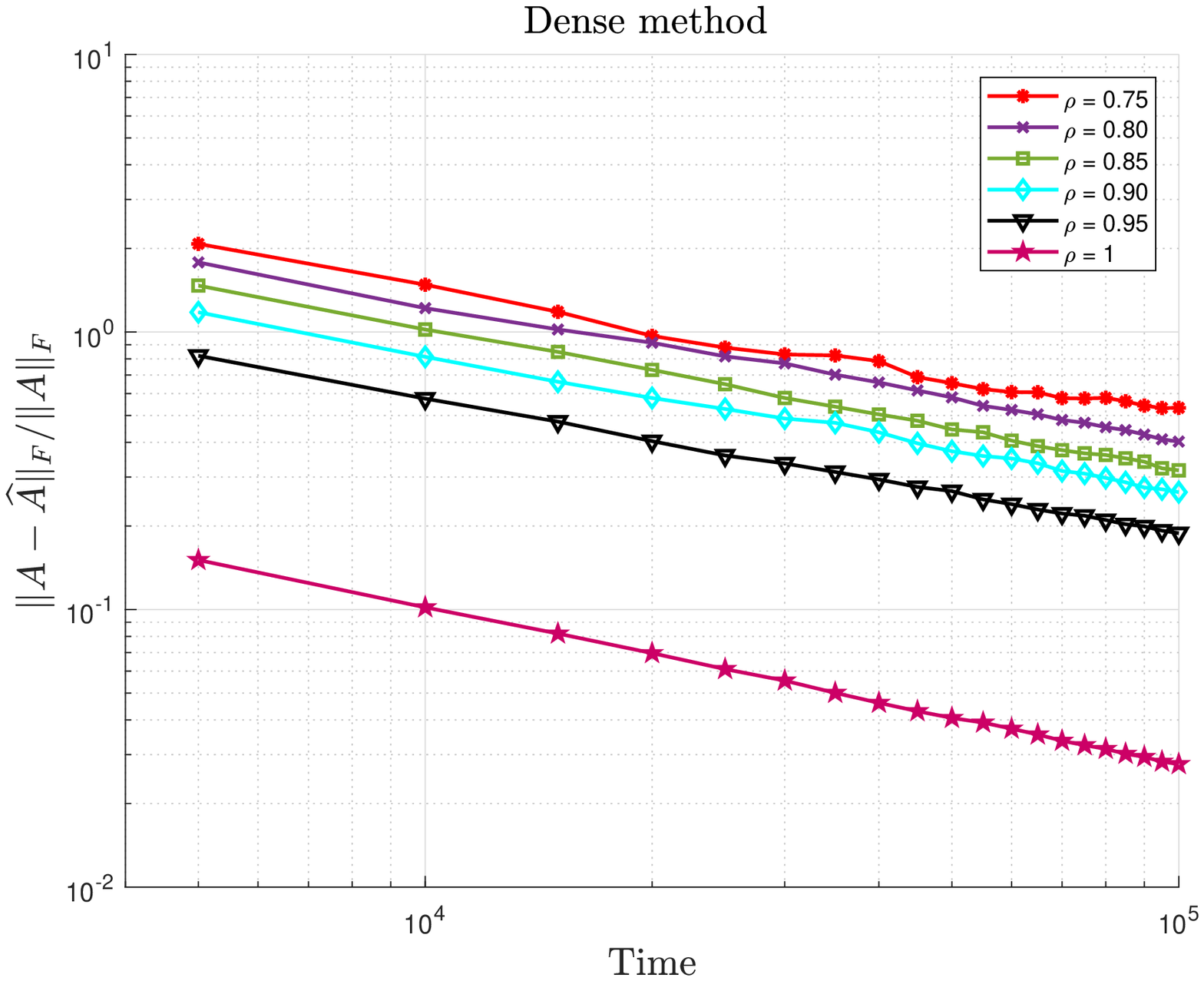} \label{fig:3a}
	}
\end{figure}
\begin{figure}[h!]
	\subfigure[Error in estimating the influence matrix with sparse method.]{
		\centering\includegraphics[width=1\columnwidth]{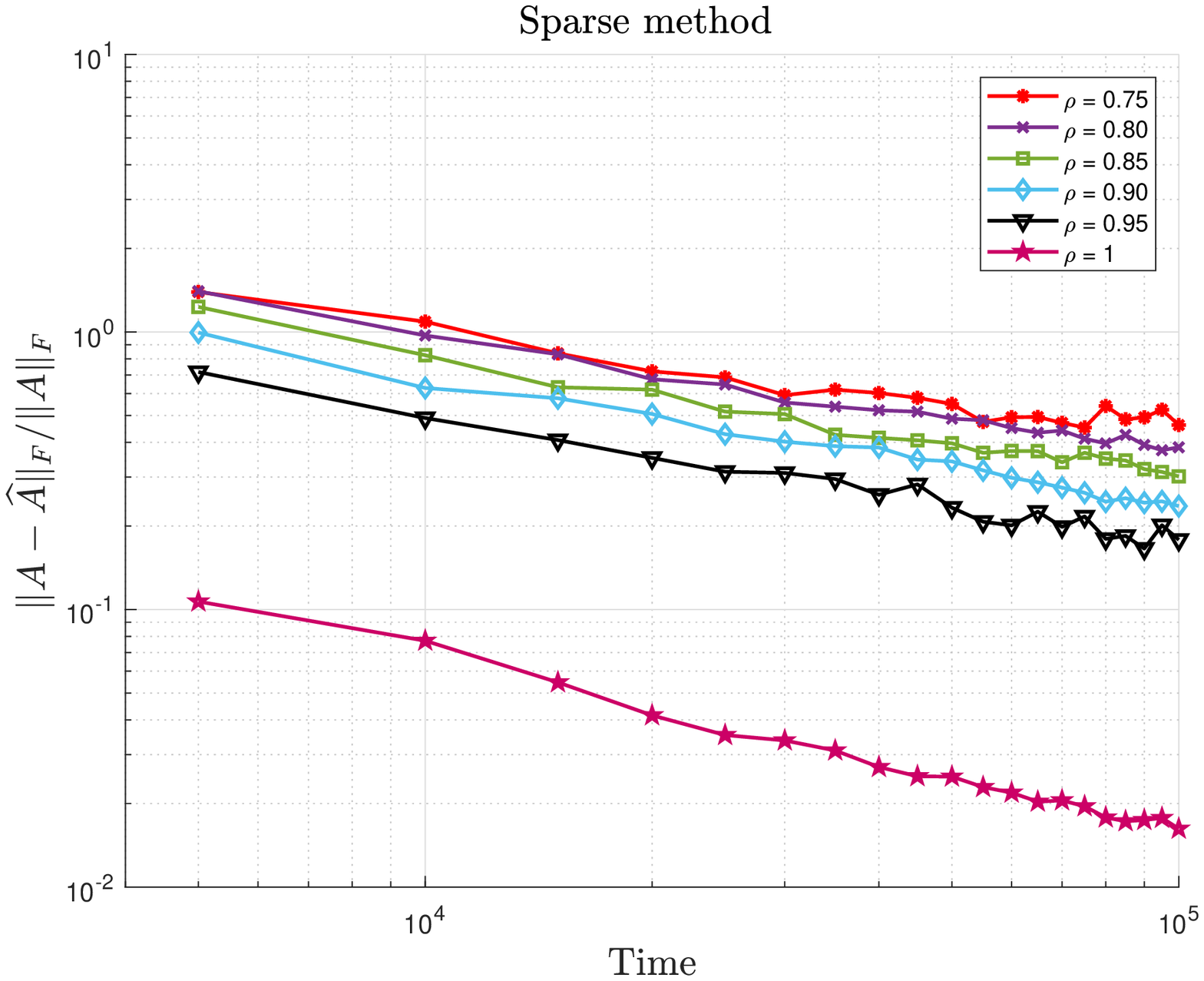} \label{fig:3b}
	}
	\caption{Comparison of dense and sparse method in a network with degree~3 using intermittent measurements with different values of $\rho$}\label{fig:3}
\end{figure}

\section{Concluding remarks}

In this paper we consider the problem of estimating the influence matrix of randomized opinion dynamics over a network from intermittent observations. Two methods are provided: i) dense method where the estimated influence matrix is directly computed from the estimated state covariance matrices and ii) sparse method where common relaxations of sparsity are used to estimate the influence matrix that has the smallest number of connections and is compatible with the information collected. Convergence of the propose methods is proven and their performance is illustrated using randomly generated gossip networks.

\begin{appendices}
\section{Proof of Theorem~\ref{thm:YW}}\label{app:A}
\begin{lemma}\label{lemma0}
Consider a substochastic matrix $M \in\R^{\V\times\V} $. If in the graph associated to $M$ there is a path from every node to a deficiency node, then $M$ is Schur stable.
\end{lemma}
\begin{lemma}\label{stability}Assume that in the graph
associated to $ W$ for any node $\ell\in \V$ there exists a path from
$\ell$ to a node $i$ such that  $\lambda_{i}<1$. Then
$$\overline{A}:=\mathbb{E}[{A(k)}]=(1-\beta)I+\beta\Lambda (I-D^{-1}(I-W))$$ and ${\overline {A^{\otimes 2}}}:=\mathbb{E}[{A(k)}\otimes A(k)]$ are Schur stable. 
Moreover,
$$
\mathrm{sr}(\overline{A})\leq 1-\beta+\beta \lambda_{\max}.$$
\end{lemma}

\textbf{Proof:}
From definition of expectation \begin{equation}\label{eq:A_bar}
\overline{A}=(1-\beta)I+\beta\Lambda (I-D^{-1}(I-W))
\end{equation}
It should be noticed that 
\begin{align*}\overline{A}_{ii}&=(1-\beta )+\beta\lambda_i(1-(1-W_{ii})/d_i)\\
\overline{A}_{ij}&=\beta\lambda_iW_{ij}/d_i\quad\text{if }i\neq j.
\end{align*}
We get $\sum_{j}\overline{A}_{ij}=(1-\beta)+\beta\lambda_i$ which is not larger than 1 if $\lambda_i<1$.
Hence, under the hypothesis, we have that $\overline{A}$ is a substochastic matrix corresponding to a graph with a path from any node $\ell$ to a node $m$ whose row sums up to less than one. By Lemma \ref{lemma0} $\overline{A}$ is Schur stable and from \eqref{eq:A_bar} we get that $\mathrm{sr}(\overline{A})\leq 1-\beta+\beta \lambda_{\max}$.

We now prove that $\overline {A^{\otimes 2}}$ is Schur stable. We observe that $\overline {A^{\otimes 2}}$ is a substochastic matrix as $\overline {A^{\otimes 2}}\mathds{1}_{|\V|^2}=\mathrm{vec}(\mathbb{E}[A(k)\mathds{1}_{|\V|}\mathds{1}_{|\V|}^{\top}A(k)^{\top}])$. Notice that
$$
[\overline {A^{\otimes 2}}]_{(i-1)|\V|+j,(h-1)|\V|+\ell}=\mathbb{E}[A_{ih}(k)A_{j\ell}(k)].
$$
Let $\widetilde{\V}=\{ (i,j):\sum_{h,\ell}\mathbb{E}[A_{ih}(k)A_{j\ell}(k)]<1.$

Since $A(k)_{ii}>0$ with probability one we have $\mathbb{E}[A_{ii}(k)A_{j\ell}(k)]>0$ iff $\mathbb{E}[A_{j\ell}]>0$ and, with the same argument we get $\mathbb{E}[A_{ih}(k)A_{jj}(k)]>0$ iff $\mathbb{E}[A_{ih}]>0$.
Given $(i_0,j_0)\in\V\times \V$, under the hypothesis, we have that there exists a sequence $(i_0,i_1,i_2,\ldots i_n,h)$ from $i_0$ to a node $h$ such that $\lambda_h<1$ and a sequence $(j_0,j_1,j_2,\ldots j_n,\ell)$ from $j_0$ to $\ell$ with $\lambda_{\ell}<1.$ It should be noticed that $(h,\ell)$ is a deficiency node for the product graph associated to $\overline {A^{\otimes 2}}$.
This implies that there exists an admissible path
$
(i_0,j_0),(i_0,j_1),\ldots,(i_0,\ell),(i_1,\ell),\ldots,(h,\ell)
$
 in the product graph associated to $\overline {A^{\otimes 2}}$.
 Using Lemma \ref{lemma0} we conclude that $\mathbb{E}[A(k)\times A(k)]$ is Schur stable.
%$$
%{\overline {A^{\otimes 2}}}=\mathbb{E}[{A(k)}\otimes A(k)]=\sum_{\theta\in\Theta}c_{\theta}(A^{\theta}\otimes A^{\theta})
%$$
%We observe that the spectrum of $\overline{A}$ is the same of $$I\otimes \overline{A}=I\otimes \sum_{\theta\in\Theta}c_\theta A^{\theta}=\sum_{\theta\in\Theta}c_\theta( I\otimes  A^{\theta})$$
%from which for every $y\in\R^{|\V|\times |\V|}$
%$$
%y^{\top}\left[I\otimes \overline{A}-{\overline {A^{\otimes 2}}}\right]y=y^{\top}\left[\sum_{\theta\in\Theta}c_{\theta}(I-A^{\theta})\otimes A^{\theta}\right]y\geq0
%$$
%where the last inequality follows from the fact that $A^{\theta}$ is positive definite and $I-A^{\theta}$ is positive semi-definite.
%We conclude that $I\otimes \overline{A}-{\overline {A^{\otimes 2}}}$ is positive semi-definite and, consequently,
%$$
%\mathrm{sr}(\overline {A^{\otimes 2}})\leq\mathrm{sr}(\overline{A})\leq (1-\beta)+\beta\lambda_{\max}.
%$$
%\end{proof}

{\bf{\em{Proof of Theorem~\ref{thm:YW}:}}}
It should be noticed that the opinions $x(k)$ are bounded, as they satisfy \begin{equation}\label{eq:bounded-states}\min_{v\in\V}{u_v}\le x_i(k)\le\max_{v\in\V}{u_v}\end{equation} for all $i\in \V$ and $k\ge0$. Then $\mathbb{E}[x(k)]$ and $\mathbb{E}[x(k)x(k)^{\top}]$ exist and are bounded.

By defining
$$
y(k)=[y_1(k)^{\top},y_2(k)^{\top},y_3(k)^{\top}]^{\top},
$$
with
$y_1(k)=\mathrm{vec}(x(k)x(k)^{\top})$, $ y_2(k)=\mathrm{vec}(x(k)u^{\top})$, $ y_3(k)=\mathrm{vec}(ux(k)^{\top}) 
$, 
\begin{align*}
Q(k):=
\left[
\begin{array}{cccc}
A(k)\otimes A(k)&A(k)\otimes B(k)&B(k)\otimes A(k)\\
0&A(k)\otimes I&0\\
0&0&I\otimes A(k)
\end{array}
\right],
\end{align*}
and $M(k)=[(B(k)\otimes B(k))^{\top},( B(k)\otimes I)^{\top},( I\otimes B(k))^{\top}]^{\top}$,
we can easily check that
\begin{equation}\label{eq:ricursion}
y(k+1)=Q(k)y(k)+M(k)\mathrm{vec}(uu^{\top})
\end{equation}
from which
\begin{align*}
\mathbb{E}[y(k+1)]&=\mathbb{E}[\mathbb{E}[y(k+1)|\mathcal{F}_k]]\\&=\mathbb{E}[Q(k)]\mathbb{E}[y(k)]
+\mathbb{E}[M(k)]\mathrm{vec}(uu^{\top})
\end{align*}
with
\begin{align*}
&\overline{Q}:=\mathbb{E}[Q(k)]\\
&=\left[
\begin{array}{ccc}
\mathbb{E}[A(k)\otimes A(k)]&\mathbb{E}[A(k)\otimes B(k)]&\mathbb{E}[B(k)\otimes A(k)]\\
0&\mathbb{E}[A(k)]\otimes I&0\\
0&0&I\otimes \mathbb{E}[A(k)]
\end{array}
\right]\\
&=\left[
\begin{array}{ccc}
\mathbb{E}[A(k)\otimes A(k)]&\mathbb{E}[A(k)\otimes B(k)]&\mathbb{E}[B(k)\otimes A(k)]\\
0&\overline{A}\otimes I&0\\
0&0&I\otimes \overline{A}
\end{array}
\right]
\end{align*}
and
$\overline{M}=\mathbb{E}[M(k)]=(\mathbb{E}[B(k)\otimes B(k)]^{\top},(\overline B\otimes I)^{\top},(I\otimes\overline B)^{\top} ]^{\top}$.
By Lemma \ref{stability} the matrix $\mathbb{E}[Q(k)]$ is Schur stable and, consequently,
$$
\lim_{k\rightarrow\infty}y(k)=(I-\overline{Q})^{-1}\overline{M}\mathrm{vec}(uu^{\top})
,
$$
from which we conclude that the sequence $y_1(k)$ and $\Sigma^{[0]}(k)=\mathbb{E}[x(k)x(k)^{\top}]$ are convergent as  $k\rightarrow\infty$.
From definition it is easy to verify that
$$
\Sigma^{[1]}(k)=\Sigma^{[0]}(k)\overline A^{\top}+\mathbb{E}[x(k)]\overline b^{\top}
$$
from which, letting $k\rightarrow\infty $ and using Proposition \ref{prop:ex_dyn}, we get that also $\Sigma^{[1]}(k) $ converges to a limit point satisfying
$$
\Sigma^{[1]}(\infty)=\Sigma^{[0]}(\infty)\overline A^{\top}+x(\infty)\overline b^{\top}.
$$

\section{Proof of Theorem~\ref{thm:performance_estimations}}\label{app:B}
Using Markov inequality \cite{} we obtain
$$
\mathbb{P}(\|\Delta x(k)\|\geq \epsilon)=\mathbb{P}(\|\Delta x(k)\|^2\geq \epsilon^2)\leq\frac{\mathbb{E}[\|\Delta x(k)\|^2]}{\epsilon^2}.
$$
We now estimate $\mathbb{E}[\|\Delta x(k)\|^2]$.
It should be noticed that by the definition of~\eqref{eq:gossip-friedkin}, the opinions $x(k)$ are bounded, as they satisfy \begin{equation}\label{eq:bounded-states}\min_{v\in\V}{u_v}\le x_i(k)\le\max_{v\in\V}{u_v}\end{equation} for all $i\in \V$ and $k\ge0$. 
As a consequence, partial observations $z(k)$ are bounded and all moments of $x(k)$ and $ z(k)$ are uniformly bounded.
Let us denote $x^{\star}=\mathbb{E}[x(\infty)]$ $e(\ell)=(z(\ell)-\pi\circ x^{\star})$ and observe that
$\pi\circ(\widehat x(k)-x^{\star})=\overline{z}-\pi\circ x^{\star}%\\
%&=
%\frac{1}{(k+1)}\sum_{\ell=0}^{k}(z(\ell)-\pi\circ x^{\star})\\
=\frac{1}{k+1}\sum_{\ell=0}^{k}e(\ell).
$
We thus have
\begin{align*}
&\Exp\| \pi\circ(\widehat x(k)-x^{\star})\|^2=\Exp\left\| \frac{1}{(k+1)} \sum_{\ell=0}^{k} e(\ell)\right\|^2\\
&=\frac{1}{(k+1)^2} \sum_{\ell=0}^{k}\Exp\left[e(\ell)^{\top} e(\ell)\right]+2 \sum_{\ell=0}^{k} \sum_{\ell=r}^{k-\ell} \Exp \left[e(\ell)^{\top} e(\ell+r)\right].
\end{align*}
From \eqref{eq:bounded-states} we can ensure that there exists a constant $\eta\in\real$ such that
$
\frac{1}{(k+1)} \sum_{\ell=0}^{k}\Exp \left[\|e(\ell)\|^2\right] \leq\eta,\ \forall k.
$
Now, it should be observed that  
\begin{align}
\nonumber
&\Exp \left[e(\ell)^\top e(\ell+r)\right]\\
\nonumber
& \quad= \Exp\left[ \Exp \left[e(\ell)^\top e(\ell+r)|P(\ell),x(\ell)\right]\right]\\
\nonumber
& \quad = \Exp\left[ e(\ell)^\top  \Exp \left[e(\ell+r)|P(\ell),x(\ell)\right]\right] \label{eq: x_l}\\
\nonumber& \quad = \Exp\left[ e(\ell)^\top   \left( \Exp \left[z(\ell+r) |P(\ell),x(\ell)\right]-\pi\circ x^{\star}  \right) \right]\\
& \quad = \Exp\left[ e(\ell)^\top    \pi\circ\left(\Exp \left[x(\ell+r) |x(\ell)\right]-x^{\star}  \right) \right].
\end{align}
By repeated conditioning on $x(\ell), x(\ell+1), \ldots, x(\ell+r-1),$ we obtain
\begin{align*}
\Exp \big[x(\ell+&r) | x(\ell) \big]=\Exp\left[A(k)\right]^{r} x(\ell)+  \sum_{s=0}^{r-1} \Exp[A(k)]^{s} \Exp[B] u\label{eq: x_l},
\end{align*}
and by recalling that $x^{\star}$ is a fixed point for the expected dynamics we get
\begin{equation}\label{eq: x_star}
x^{\star}=\Exp\left[A(k)\right]^{r} x^{\star}+  \sum_{s=0}^{r-1} \Exp[A(k)]^{s} \Exp[B(k)] u.
\end{equation}
From equations \eqref{eq: x_l} and \eqref{eq: x_star} we obtain
\begin{align*}
&\Exp \left[e(\ell)^\top e(\ell+r)\right]\\
&\quad= \Exp\left[e(\ell)^\top  \pi\circ(\Exp\left[A(k)\right]^{r} (x(\ell)-x^{\star}))\right]\\
&\quad= \Exp\left[(P(\ell)x(\ell)-\pi\circ x^{\star})^\top  \pi\circ(\Exp\left[A(k)\right]^{r} (x(\ell)-x^{\star}))\right]\\
&\quad= \Exp\left[\pi\circ (x(\ell)-x^{\star})^\top  \pi\circ(\overline{A}^{r} (x(\ell)-x^{\star}))\right]\\
&\quad\leq{\eta}\nu^r,
\end{align*}
where, by Lemma~\ref{stability}, $\nu=\mathrm{sr}(\overline A)<1$.
Finally, we have
\begin{align*}
\Exp\left[\| \pi\circ{(\widehat x(k)-x^{\star})}\|^2\right]&\leq\frac{\eta}{(k+1)^2}\left(k+1+2\sum_{\ell=0}^{k-1}\sum_{r=0}^{k-\ell}\nu^r\right)\\
&\leq\frac{\eta}{(k+1)}\left(1+\frac{2}{1-\nu}\right).
\end{align*}
Since $\nu<1$ we have there exists $C>0$ such that
\begin{align*}
\Exp\left[\| \pi\circ{(\widehat x(k)-x^{\star})}\|^2\right]&\leq\frac{1}{(k+1)}\frac{C}{1-\nu}
\end{align*}
and, consequently, we get
\begin{align*}
\Exp\left[\| {\widehat x(k)-x^{\star}}\|^2\right]&\leq\frac{C}{(k+1)(1-\nu)(\pi^{\star})^2},
\end{align*}
where $\pi^{\star}=\min_{v\in\V}\pi_v$.

The proof of the second part of theorem follows the same arguments using the recursion in \eqref{eq:ricursion} and we omit for brevity. 
\end{appendices}

\bibliographystyle{IEEEtran}
\bibliography{CDC18_biblio.bib}

\end{document}